\documentclass{article}
\usepackage{amsmath,graphicx,cite}
\begin{document}

\author{Allison N. Dickey \and
Roland Faller
\footnote{to whom correspondence should be addressed at rfaller@ucdavis.edu}\\
\small{Department of Chemical Engineering \& Materials Science,}\\
\small{UC Davis, Davis, CA 95616, USA}
}
\title{Investigating Interactions of Biomembranes and Alcohols: A
Multiscale Approach}

\maketitle
\begin{abstract}
We study the interaction of lipid bilayers with short chain
alcohols using molecular dynamics on different length scales. We
use detailed atomistic modeling and modeling on the length scale
where an alcohol is just an amphiphilic dimer. Our strategy is to
calibrate a coarse--grained model against the detailed model at
selected state points at low alcohol concentration and then
perform a wider range of simulations using the coarse--grained
model. We get semiquantitative agreement with experiment for the
major observables such as order parameter and area per molecule.
We find a linear increase of area per molecule with alcohol
concentration. The alcohol molecules in both system descriptions
are in close contact with the glycerol backbone. Butanol molecules
can enter the bilayer to some extent in contrast to the behavior
of shorter alcohols. At very high alcohol concentrations we find
clearly increased interdigitation between leaflets.
\end{abstract}
\section{Introduction}
Phospholipid bilayers serve an important role in all living cells
since they constitute the bulk of cellular and intracellular
membranes. Therefore, it is not surprising that there is significant
attention and research devoted to the exploration of phospholipid
behavior. Computer simulations play an ever increasing role in this
field. Detailed computer simulations of phospholipid monolayers and
bilayers have recently achieved a high degree of
sophistication~\cite{feller95,tieleman97,tobias97,bandyopadhyay98,husslein98,rog01,saiz02a,patra03,patra04,leontiadou04}.
The interactions between lipid bilayers and water have been thoroughly
examined and these investigations have resulted in refined bilayer
structural
models~\cite{husslein98,mashl01,saiz02a,gurtovenko04}. Mixtures of
various phospholipids~\cite{pandit03,balali03,vries04,gurtovenko04} as
well as of phospholipids and cholesterol
systems~\cite{tu98,smondryev00,pandit04,falck04} have been recently
characterized. However, the interactions between lipid bilayer
membranes and small molecules have not gained much attention except
for a few initial studies with
alcohols~\cite{feller02,bemporad04,patra04s},
sugars~\cite{sum03a,pereira04}, and dimethylsulfoxide
(DMSO)~\cite{smondryev99,sum03b}.

Since cell membranes are the first part of the cell to contact
with any nutrient or pathogen in the extracellular matrix, it is
extremely important to fully understand membrane interactions with
biologically relevant small molecules.

It has been experimentally observed that alcohols have a
destabilizing effect on model membranes~\cite{ly02,ly04} by
causing an increase in area per molecule and a decrease in bilayer
thickness. This alcohol induced change in bilayer structure is
directly related to the industrial problem of stuck fermentation
in the wine industry~\cite{bisson02,cramer02}. A stuck
fermentation occurs when yeast cells do not transform all
available sugar into alcohol. It has been proposed that the
underlying mechanism of this problem is an alcohol--triggered
structural transition in the membrane. This results in a
conformational change in trans--membrane proteins and causes the
proteins to become dysfunctional~\cite{bisson02}. To date, there
is no available method for predicting a stuck fermentation. By
examining the influence of alcohols, such as butanol, on lipid
bilayers, we will be able to study the primary mechanisms that
cause stuck fermentations, as protein conformational changes are
predicted to be a secondary effect.

To determine how the entire bilayer area is affected by
interactions with n--butanol molecules, a multiscale approach is
used. Coarse--grained simulations can be used to study large scale
fluctuations and membrane curvature (on the order of tens of
nanometers). Detailed atomistic simulations on the scale of a few
angstroms are appropriate for studying interactions between
neighboring atoms. Surprisingly, biomembrane multiscale models are
not as abundant as polymer multiscale models. There have been a
few proposed coarse--grained
models~\cite{smit90,goetz99,mouritsen00,soddemann01,shelley01,ayton02,guo03,mueller03,kranenburg03,marrink04}
and they have the advantage of being economical in computer power
because a number of atoms or even whole molecules are combined
into fictitious units and are assigned an interaction potential.
This potential is often designed to be computationally efficient
rather than an accurate system representation. However, generic
chemical effects such as hydrophilic--hydrophobic
interactions~\cite{soddemann01}, or the anisotropy of the overall
molecule~\cite{mouritsen00} are usually included in the potential.
Even though the atomistic detail is missing, interesting generic
properties of membranes have been elucidated with these models.
One example is the general pathway of self--assembly, which is not
specific to the individual
lipid~\cite{mouritsen00,soddemann01,shelley01}.

The purpose of this contribution is to study the effects of low
molecular weight alcohols, such as n--butanol, by combining a fully
atomistic model with a recently proposed coarse--grained
model~\cite{marrink04,faller04c}.

\section{Simulation Techniques}
We used molecular dynamics simulations to investigate the interactions between phospholipid bilayers and n--butanol at
varying alcohol concentrations. Both an atomistic--scale and
coarse--grained model were used. All simulations were performed at 325~K and
atmospheric pressure (1~atm). The simulated membranes consisted of fully
hydrated dipalmitoylphosphatidylcholine (DPPC) bilayers with 128
molecules, i.e., 64 per leaflet. DPPC was chosen because it is one of
the most abundant phospholipids in animal cell membranes. We performed
our simulations for both scales using the GROMACS
simulation suite~\cite{gromacs01}.
\subsection{Atomistic Modeling}
Our detailed simulations are fully atomistic with an exception for
hydrogen atoms that are bonded in methyl(ene) groups both in
butanol as well as in the lipid tails. These hydrogens together
with the respective carbon are collapsed into one united atom
description. Simulations of the pure bilayer, as well as bilayers
with up to 5 wt\% n--butanol, (lipid free basis, as we do not take
the lipids into account for concentration calculations) have been
performed. Simulations were performed under constant temperature
and constant pressure conditions using the Berendsen
weak--coupling scheme~\cite{berendsen84}. The coupling times were
$\tau_p=1.0$~ps and $\tau_T=0.2$~ps for pressure and temperature
respectively and a compressibility of
$1.12\times10^{-6}$~atm$^{-1}$. The simulations with a time--step
of 2~fs lasted up to 10~ns. We used a cutoff of the Lennard--Jones
interaction of 1.0~nm.  The lipid simulation models were designed
by Berger {\it et al.}~\cite{berger97}. Additionally, we use the
Gromos model for alcohols~\cite{gromacs01} combined with the SPC/E
model~\cite{berendsen87} for water. Electrostatic interactions
have been considered using the particle mesh Ewald
technique~\cite{essmann99}. The initial system configurations were
taken from earlier studies on shorter alcohols~\cite{lee04} where
the ethanol was replaced by n--butanol.
\subsection{Coarse--Grained Modeling}
For the coarse--grained simulations our water and lipid
interaction potential parameters come from a model proposed by
Marrink {\it et al}\cite{marrink04}\cite{marrink04}. The initial
system configurations were deduced from an earlier study on pure
DPPC~\cite{marrink04}\footnote{Model and configurations available
for download at http://md.chem.rug.nl/~marrink/coarsegrain.html}.
The model was originally parameterized to reproduce the
structural, dynamic, and elastic properties of both lamellar and
non--lamellar phospholipid states. Groups of 4--6 heavy atoms are
combined into coarse--grained interaction sites and are classified
according to their hydrophobicity. The lipid headgroup consists of
four sites. There are two hydrophilic sites: one representing the
choline and one representing the phosphate group, and two
intermediately hydrophilic sites capable of hydrogen bonding
representing the glycerols. Each of the two DPPC tails is modeled
by four hydrophobic sites. Water is represented by hydrophilic
interaction sites, where each site represents four real water
molecules. All sites interact in a pairwise manner via a
Lennard--Jones (LJ) potential.  Five different LJ potentials are
used and range from weak for hydrophobic interactions to strong
for hydrophilic interactions.

In addition to the LJ interactions, a screened Coulomb interaction is
used to model the electrostatic interaction between the zwitterionic
headgroups. The choline group bears a charge of $+1$, and the phosphate
group bears a charge of $-1$.  Soft springs between bonded pairs keep
the molecules intact. Angle potentials provide the appropriate
chain stiffness and correct conformation.
For efficiency reasons all CG atoms have the same mass of 72 atomic units in
the simulation.

For our studies, we had to devise a model for the alcohol as it
was not defined in the original force--field. We use the strategy
explained in the manual~\cite{marrinkmanual}. Thus, our alcohols
are modeled as a dimer of a hydrophilic site and a hydrophobic
site. The hydrophilic site interacts like the water molecules, the
hydrophobic site is the same as the alkanes in the lipids. We
performed additional simulations where the hydrophilic site was
exchanged against an interaction like the one of the glycerols,
this did not change the conclusions significantly.  We are aware
that this model makes the alcohol a symmetric amphiphile which is
not fully realistic. As the interaction centers in the lipid tails
stand for four carbons we use this coarse--grained alcohol as
n--butanol and refer to it that way in the remainder of the
article. Note, that we did not do any re--optimization of any
parameter. We experimented with slightly different
parameterizations but these were not as successful. This may be
attempted in subsequent work but here we want to check the quality
of the proposed model.

All simulations are set up in the bilayer state and we do not observe
any tendency of instability of the bilayer on the timescale of the
simulations which lasted for up to 2.4$\mu$s. Note, that this is an
effective time. A scaling factor of four was previously found to
reproduce both lipid lateral diffusion rates and the self--diffusion
of water for the CG model~\cite{marrink04}. The times reported in this
paper will therefore be effective times which are physically
meaningful. For example we use a time--step of 40~fs which is
considered to represent 160~fs.

The box dimensions are again coupled semi--isotropically to a pressure
bath of 1~atm~\cite{berendsen84}. The temperature of the system is
controlled using a weak coupling scheme~\cite{berendsen84} with
coupling times $\tau_p=1$~ps, $\tau_T=10$~ps, measured in the rescaled
time scale. The compressibility was set to
$5\times10^{-6}$~atm$^{-1}$. The cutoff for Lennard--Jones
interactions as well as the electrostatics was $R_C=1.2$nm.

\begin{figure}
\includegraphics[height=4cm]{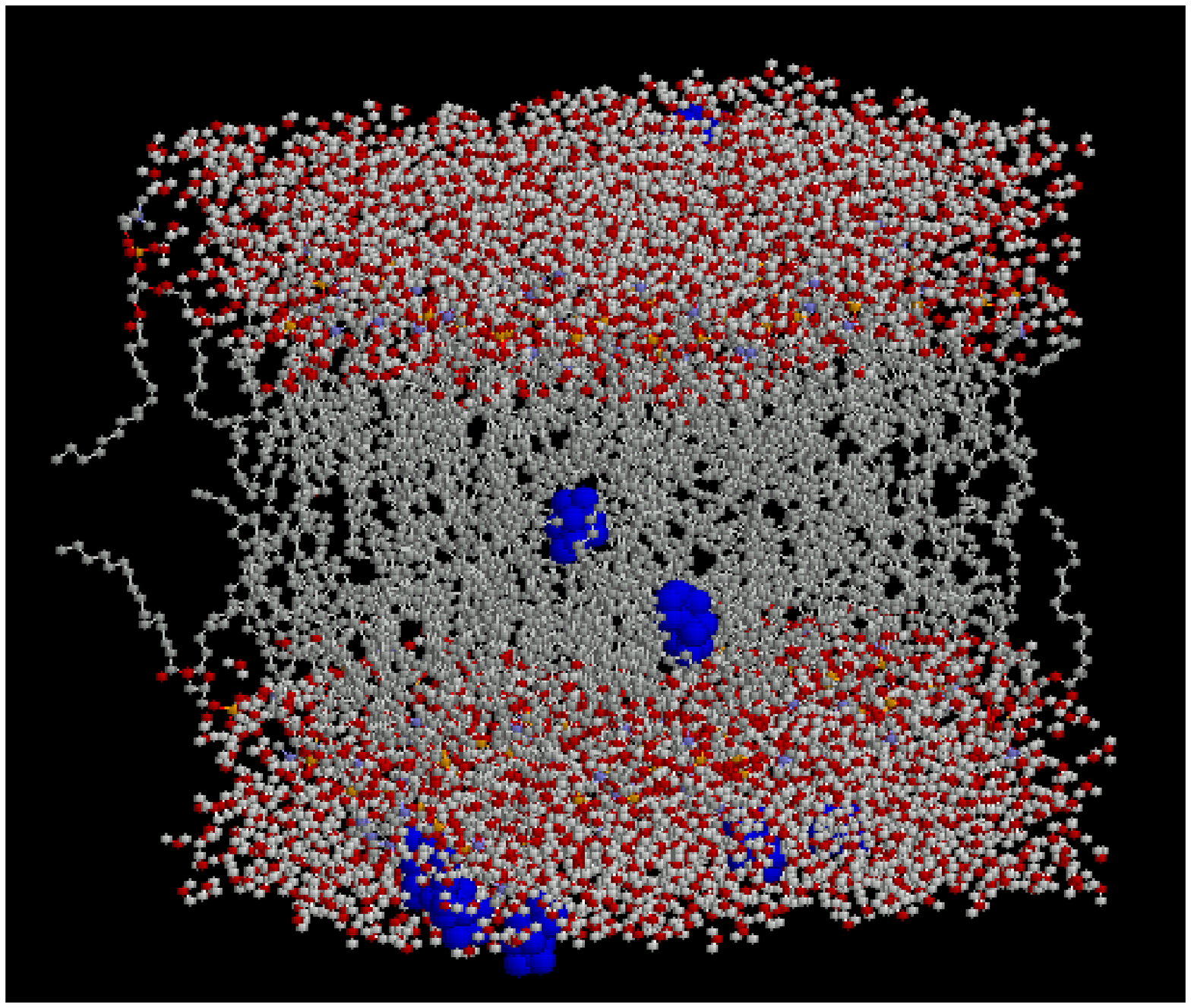}
\includegraphics[height=4cm]{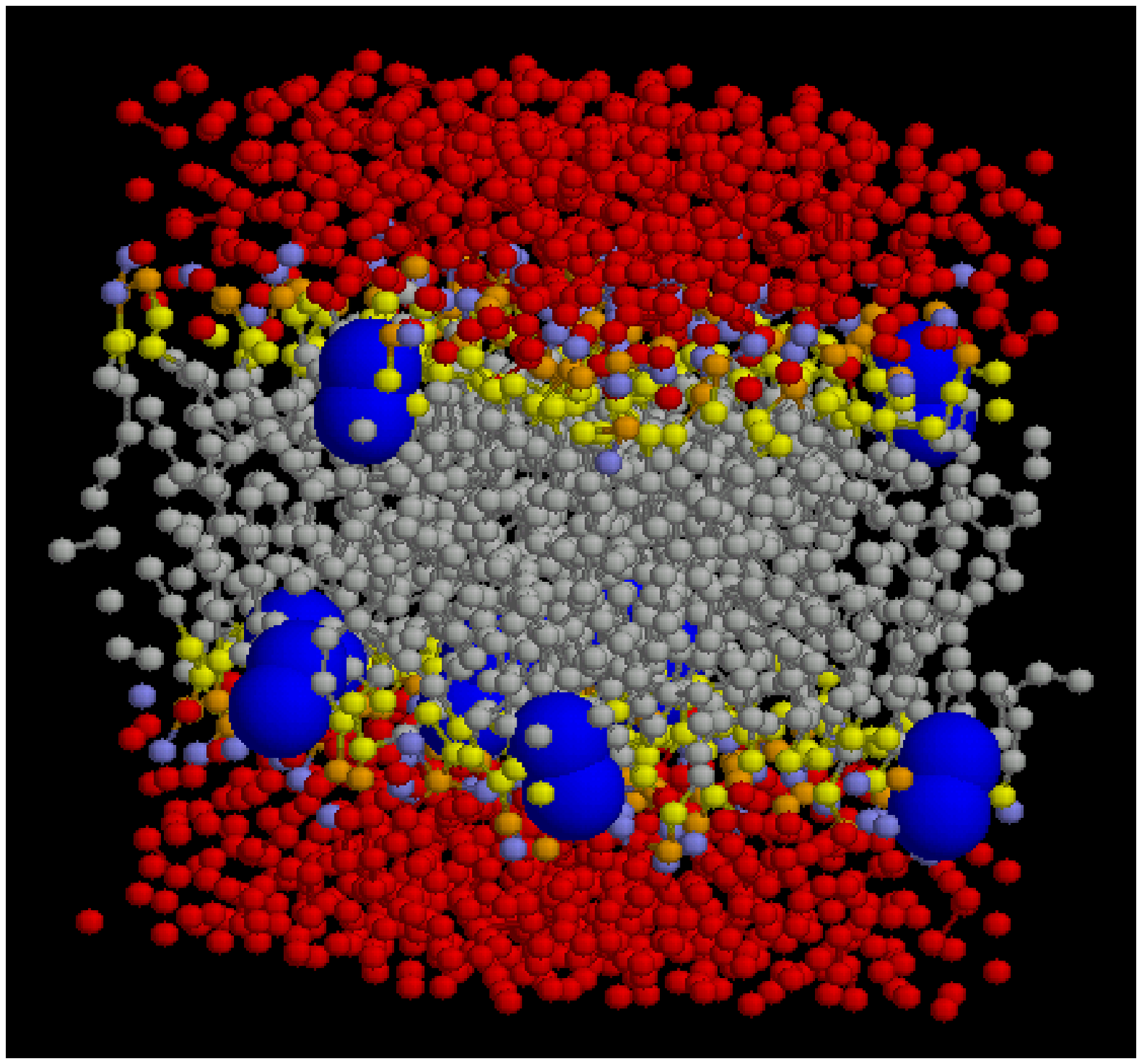}
\caption{Snapshots: Left: atomistic system with 8 butanols, Right:
coarse--grained system with 10 butanols. The two systems are very
close to each other in (rescaled) concentration. Butanol molecules
are highlighted in blue and increased in size for clarity. Note,
that the water in the coarse--grained model are red due to the
oxygens.} \label{fig:snap}
\end{figure}

The water rescaling is also visible in the alcohol concentrations.
Since one CG water represents four water molecules whereas one CG
butanol represents one real n--butanol, so that a simulated
concentration of 1:100 (butanol : CG water) actually represents
1:400. This renormalization was taken in consistency with the
water diffusion, no further change of parameters was applied.
Table~\ref{tab:sys} summarizes all simulations considered in this
work and Figure~\ref{fig:snap} shows snapshots of an atomistic and
a coarse--grained simulation in comparison.
\begin{table}
\begin{tabular}{crrl}
System Type & \# Alcohols & \# Waters & molar concentration \\
CG & 0 & 1300 & 0.0\\
CG & 5 & 1300 & 0.0010\\
CG & 7 & 1300 & 0.0013\\
CG & 10 & 1300 & 0.0019\\
CG & 15 & 1300 & 0.0028\\
CG & 25 & 1300 & 0.0048\\
CG & 50 & 1300 & 0.0095\\
CG & 75 & 1300 & 0.0142\\
CG & 100 & 1300 & 0.0189\\
ATOM & 0 & 3655 & 0.0\\
ATOM & 8 & 3655& 0.0021\\
ATOM & 39 & 3655& 0.0105\\
\end{tabular}
\caption{Overview of the simulated systems. ATOM denotes simulations in atomistic detail and CG
represents coarse--grained
simulations. Note that in the case of coarse--grained simulations the number
of coarse grained waters corresponds to 4 times as many real water molecules
and is considered correspondingly in the concentrations}
\label{tab:sys}
\end{table}

\section{Results and Discussion}
The most important use of a coarse--grained model is that
significantly longer time scales can be achieved with this model
than if an atomistic model were applied. In our case
coarse--grained simulations which were running for 600~ns
simulation time (2.4$\mu$s effective time) took 40 hours on a AMD
Athlon. On the same computer system the atomistic simulations take
about 1.5 days for a nanosecond so the speedup is around three
orders of magnitude. However, it is important that the
coarse--grained model appropriately represents the underlying
system. To this end we are comparing a number of important
variables. The area per molecule is one of the easiest
experimental observables for such a system. It has been measured
for n--butanol and other low molecular weight
alcohols~\cite{ly02,ly04}. The experimental system is not exactly
the same as our simulations as it was a SOPC bilayer at T=298~K.
SOPC is unlike DPPC not fully saturated and both tails are 4
carbons longer, such that quantitative comparisons cannot be made.
In the simulations we measure the area per molecule as the area of
the system in the $xy$--plane -- the $z$ direction is the bilayer
normal -- divided by the number of lipids per leaflet (here 64).

\begin{figure}
\includegraphics[width=7cm]{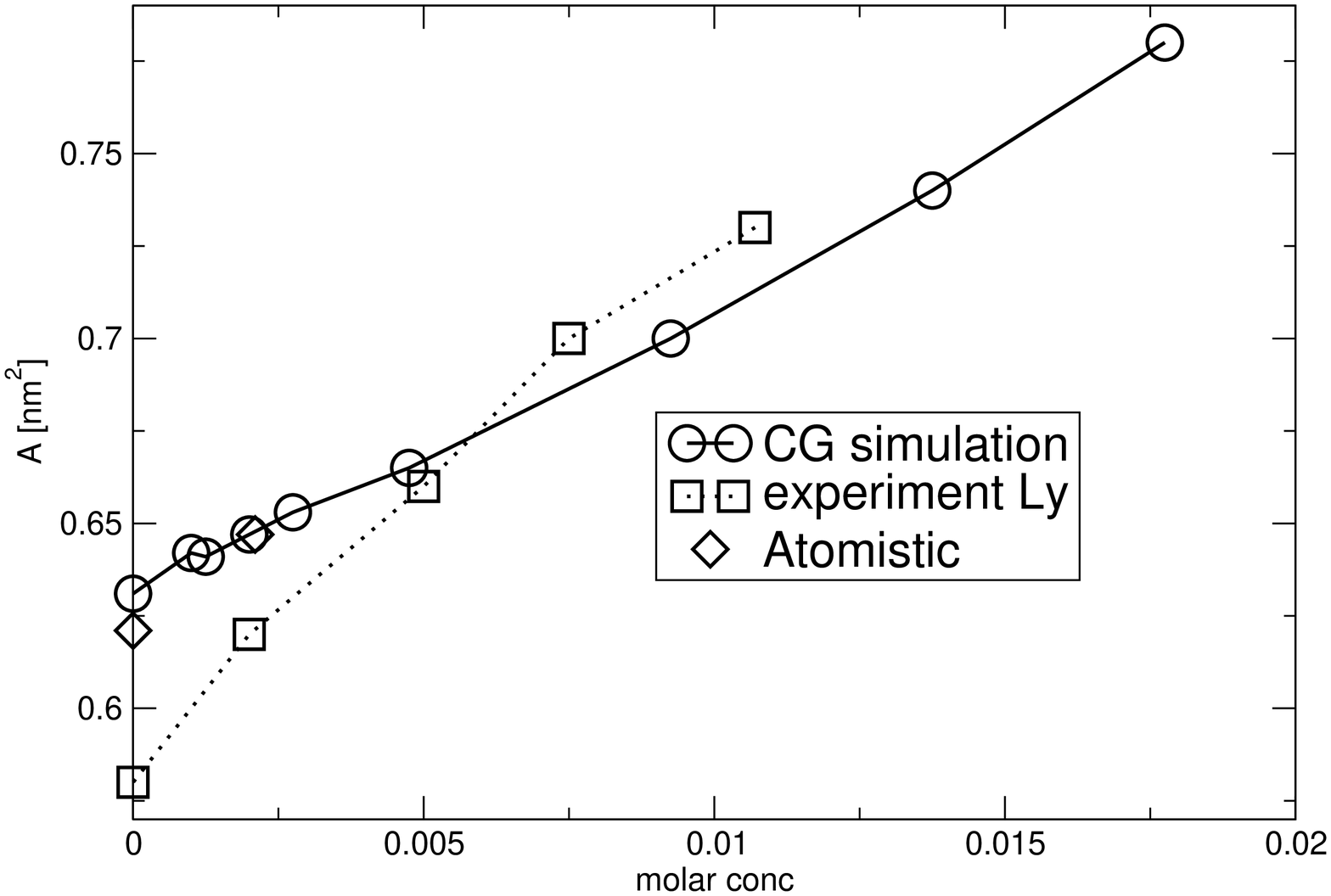}
\caption{Area per molecule depending on butanol concentration comparing
the atomistic and the coarse--grained model with experiments. Lines
are meant as guide to the eye only. Note, that in the coarse grained
simulations the concentrations are not the direct concentrations but
renormalized according to the meaning of the water molecules (see text
for details). The experimental data from ref.~\cite{ly04} are measured
using another lipid (SOPC instead of DPPC). We estimate the errors in
the atomistic simulations to be about 5\% and in the coarse grained
case to be about 2\%.}
\label{fig:area}
\end{figure}

Figure~\ref{fig:area} shows the dependence of the area per molecule on
alcohol concentration, we also show data from the experiments of Ly et
al.~\cite{ly04} in that figure. We obtain good agreement between all
three data sets. Both simulation models as well as the experimental
data exhibit a linear increase of the area per molecule with alcohol
concentration. The agreement between the atomistic and the
coarse--grained simulations is satisfactory. Only the coarse--grained
simulations allow us to access to high alcohol concentrations due to
the steep increase in equilibration times. The agreement between
experiments and the coarse--grained model is reasonable considering
the difference in systems. Reoptimization of the alcohol parameters
both of the atomistic as well as the coarse--grained model will be
made in order to reproduce experiments more exactly. The lipid
parameters on the other hand are very appropriate for our studies. If
we compare to experiments on pure DPPC in the bilayer state we are in
very good agreement with the typical value of
63\AA$^2$~\cite{nagle02}. From our data we obtain an area expansion
coefficient by linear regression of $\kappa=14$ nm$^2$ for the
coarse--grained model in comparison to $\kappa=8$ nm$^2$ for the
experiments. As we are only able to equilibrate low alcohol
concentrations in the atomistic model we refrain from estimating this
coefficient for the detailed model. This number measures the
extrapolated difference between a bilayer in pure water against one in
pure alcohol. This semiquantitative agreement between experiments and
the coarse--grained model is in line with the recent conclusions that
the same model can reproduce the phase coexistence in a mixed bilayer
but the phase transition temperature is underestimated~\cite{faller04c}.

\begin{figure}
\includegraphics[height=4cm]{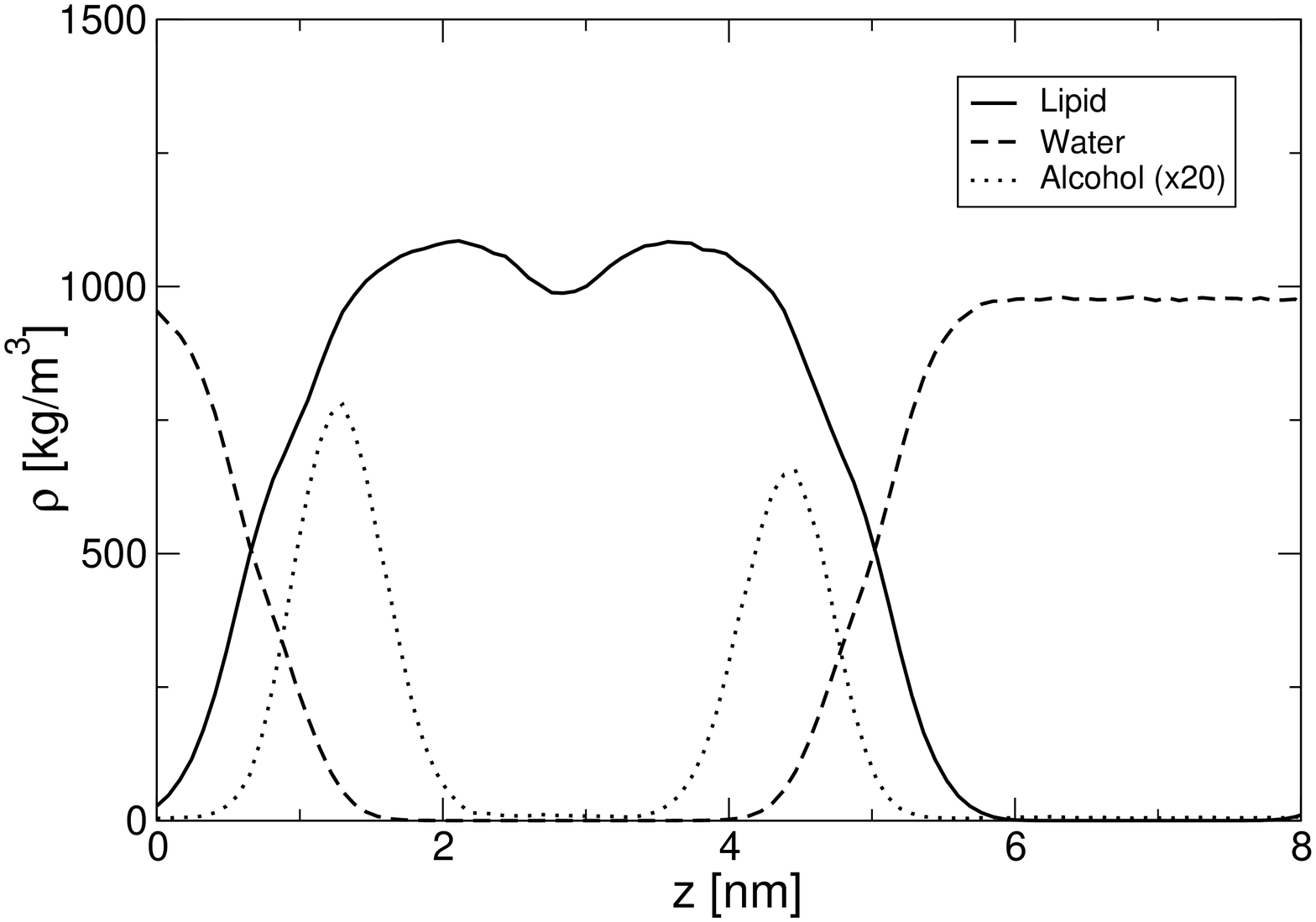}\hspace{1cm}
\includegraphics[height=4cm]{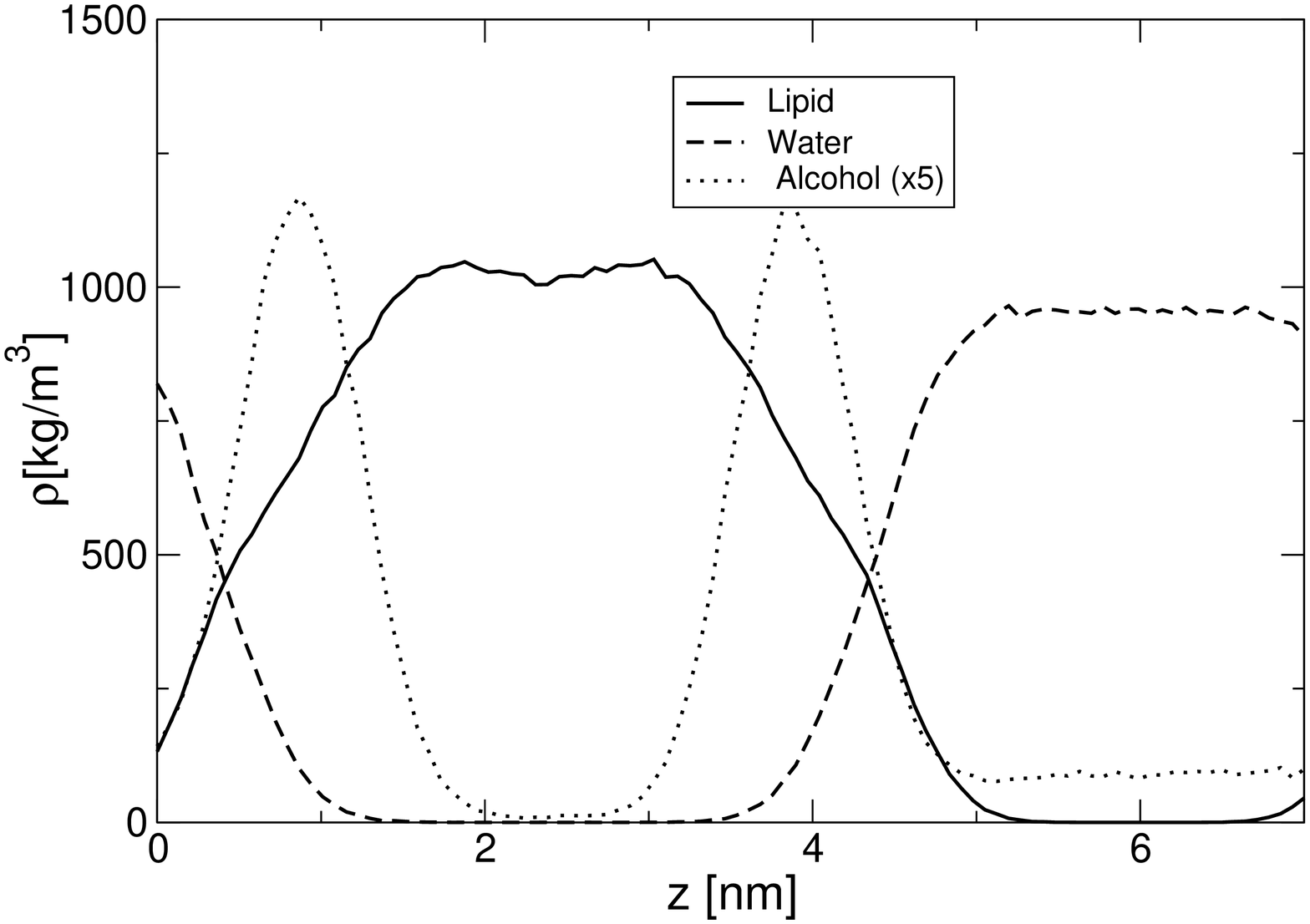}\\

\includegraphics[width=5.2cm]{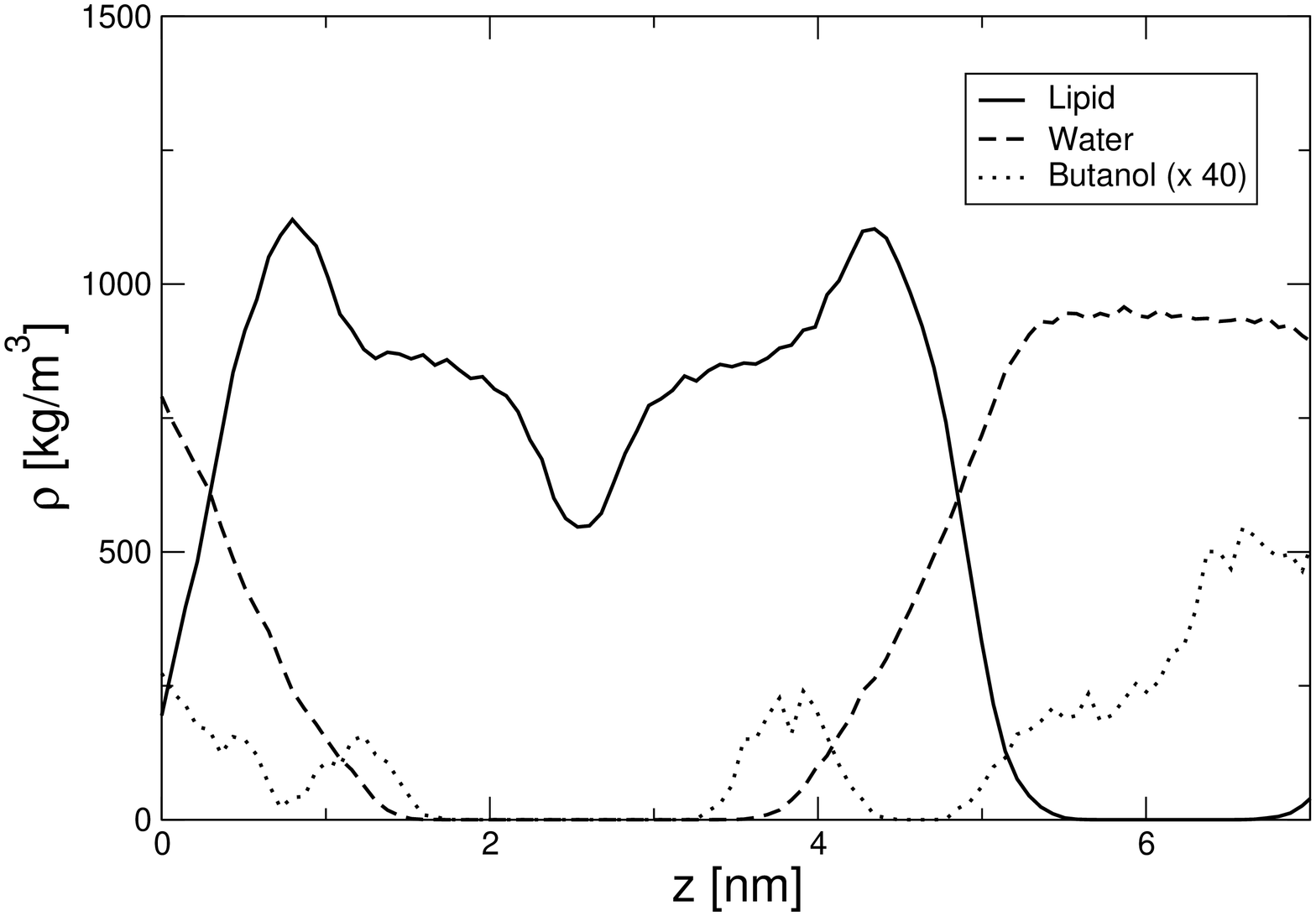}
\caption{Density profile in coarse--grained and atomistic simulations.
The upper panels represent coarse--grained simulations including 10
(left) or 100 (right) butanol molecules corresponding to molar
concentrations of 0.0019 and 0.0189, respectively. The lower panel
represents the atomistic data at a concentration of 0.0105.
Note that for clarity the butanol concentration is multiplied by different
factors as indicated in the figures.}
\label{fig:densprof}
\end{figure}
It is also of interest to determine where the butanol molecules
are preferentially located within our system. This is deduced
through density profiles which show a clear separation in a
bilayer and a water phase. The headgroup are has the highest
density and the center plane of the bilayer, which is its symmetry
plane, is the plane of lowest density. This profile agrees very
well with earlier atomistic and coarse--grained
simulations~\cite{tieleman97,marrink04}. The butanol molecules are
mainly located at the interface.This has previously been reported
for ethanol and methanol~\cite{feller02,patra04s}. It is
interesting to note that the butanol concentration in the bilayer
interior is not zero. Atomistic detailed simulations of shorter
alcohols have shown that methanol does not exist in the
bilayer~\cite{patra04s} and that ethanol was able to cross the
bilayer but could not reside in it~\cite{patra04s}. Generally, the
probability of bilayer penetration increases with alcohol chain
length due to the increasing hydrophobicity.

Comparing the coarse--grained with the atomistic model reveals
some differences. First, it is clear that the atomistic density
profile exhibits more details than can be represented in a
coarse--grained model.

\begin{figure}
\includegraphics[width=5.5cm]{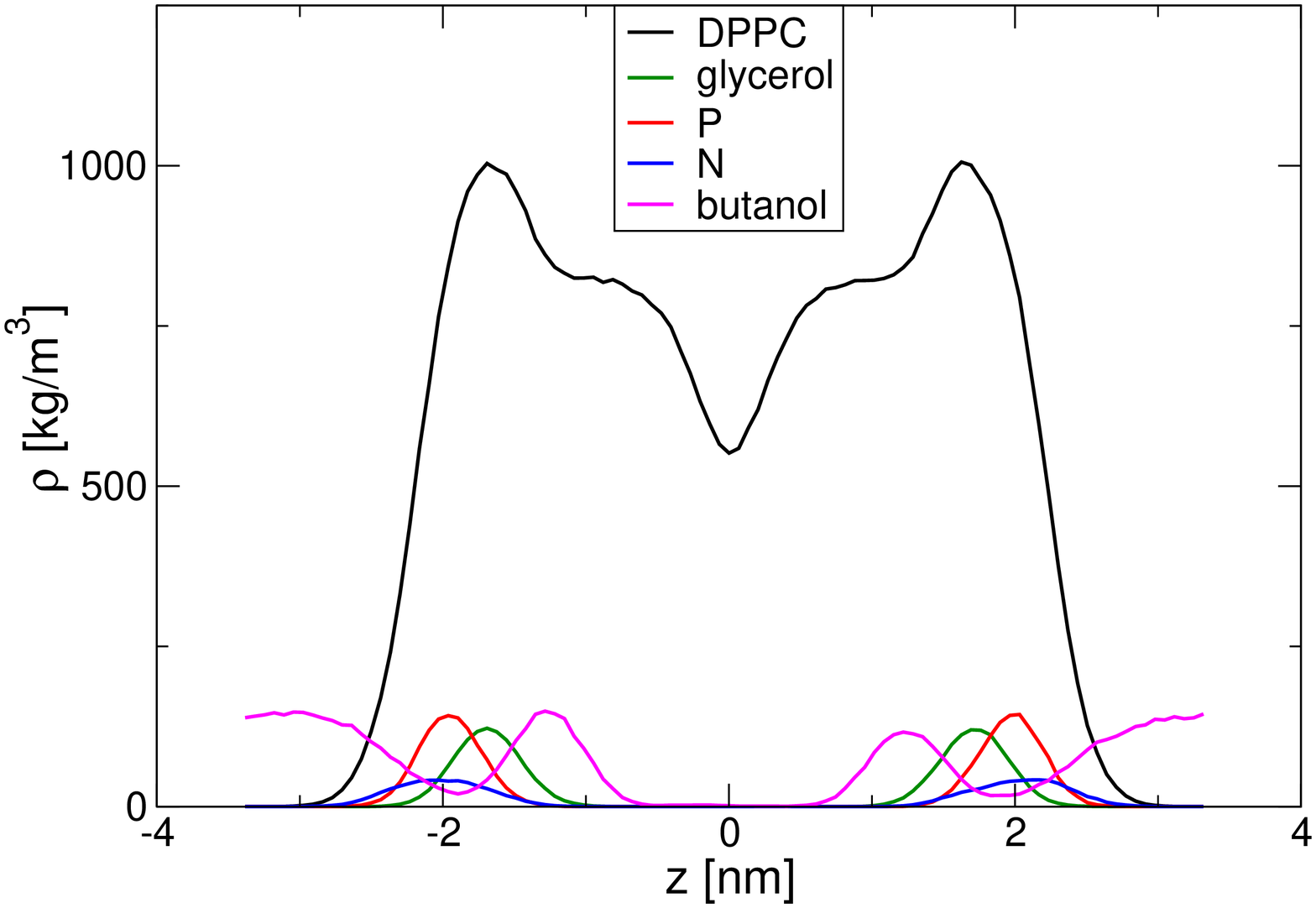}\hspace{1cm}
\includegraphics[width=5.5cm]{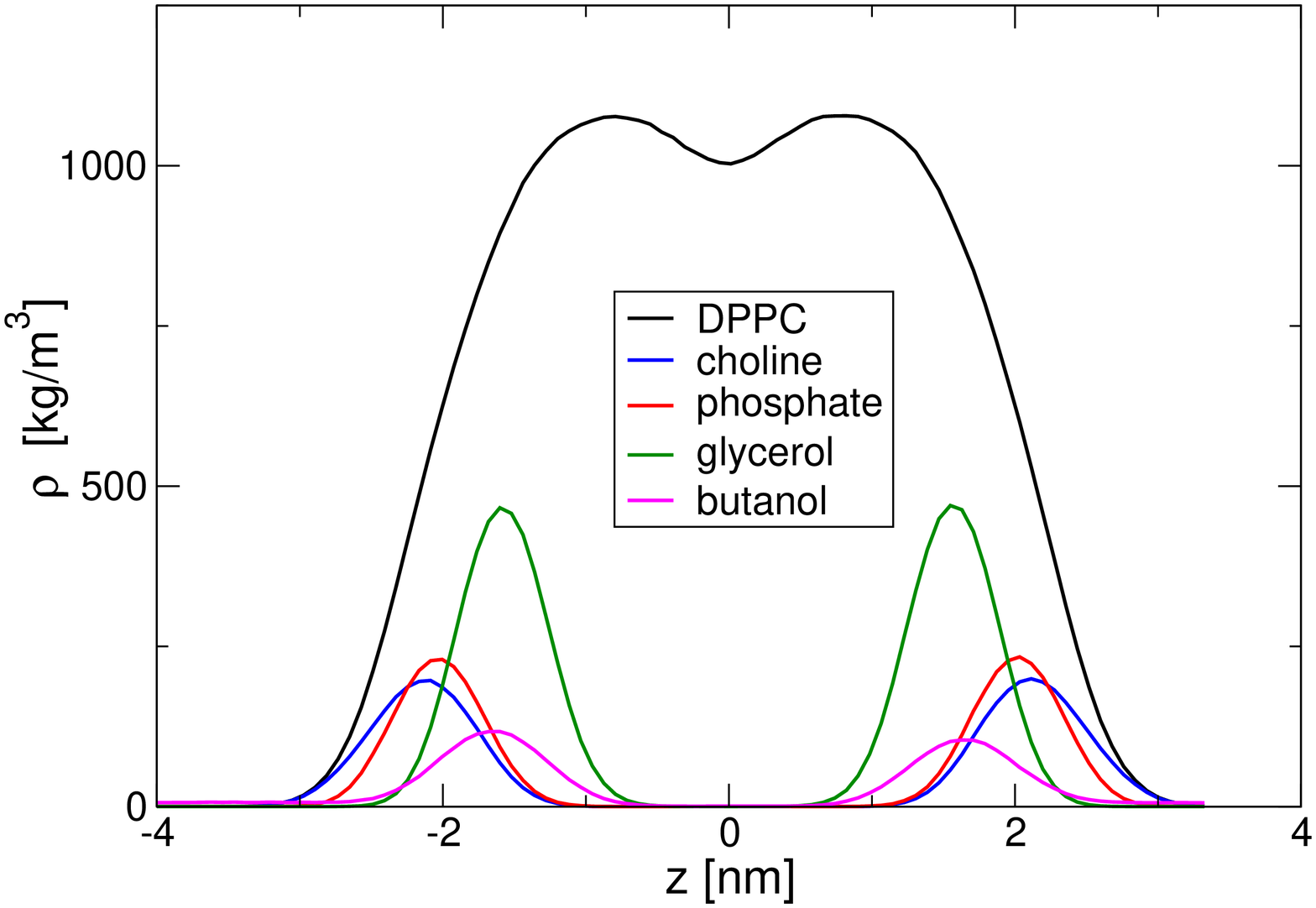}

\caption{Density profiles of the bilayer, resolved according to the
various groups in the lipids. Left: atomistic (concentration 0.0021),
Right: coarse--grained (concentration 0.0019). Note that due to the
equal weights of the coarse--grained groups the relative heights of
the peaks cannot be directly compared.}
\label{fig:densprofresol}
\end{figure}

However, a more important difference between the models can be
seen in the concentration of butanol in the water phase. The
coarse--grained model overestimates the affinity of alcohols to
the phase boundary. In contrast to the atomistic simulations the
butanol concentration in the coarse--grained model vanishes almost
in the water.  Currently, the length of the atomistic simulation
is too short to fully determine the extent of butanol penetration
in the bilayer. In general, the coarse grained model can be used
to make a basic sketch of the overall system features. However,
future refinement is desirable. Figure~\ref{fig:densprofresol}
allows a more detailed investigation of the butanol position as we
resolve the positions of the various sub-groups of the
phospholipids along the bilayer normal. We see in both models that
butanol stays in close contact to the lipid glycerol moiety. This
is in agreement with earlier studies on shorter alcohols showing a
strong hydrogen bonding of the alcohols to the oxygens in the
glycerol~\cite{feller02,patra04s}. In the atomistic model we
actually find the alcohol residing between the glycerol and the
alkane tails whereas in our coarse--grained model the butanol and
glycerol layer coincide.  Another important structural quantity
where we obtain excellent agreement is the overall bilayer
thickness.  We use the distance between the phosphorus (or
phosphate) planes as an indicator of thickness which leads to
3.96~nm in the atomistic case and 4.06~nm for the larger scale
model. The latter number actually does not change at all if we use
a butanol model where we exchange the water--like interaction of
the OH group with an interaction site like the glycerols in the
lipid.

\begin{figure}
\begin{center}
\includegraphics[width=7cm]{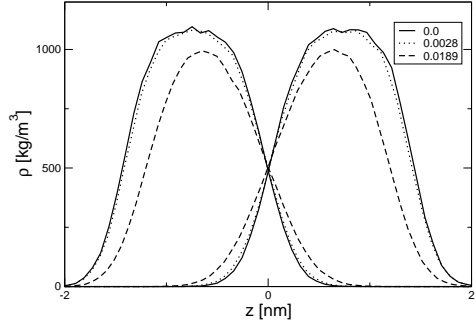}
\caption{Density profiles for the alkane tails of the lipids from
the coarse--grained simulations. The two leaflets are studied
separately to examine the likelihood of interdigitation. All
curves are shifted that the plane of equal density between the
leaflets (symmetry plane of the bilayer) is at $z=0$.}
\label{fig:interdig}
\end{center}
\end{figure}

If we compare the two concentrations of the coarse--grained model
shown in Figure~\ref{fig:densprof} we see that at the higher
concentration the bilayer configuration becomes distorted at the
interface. This occurs because the native interface cannot
accommodate all alcohols. Additionally, the plane of lowest
density is not as pronounced as it is with the lower
concentration. This corresponds to experimental and simulational
suggestions of increasing interdigitation between layers with
alcohol concentration~\cite{mou94,adachi95,kranenburg03}. In order
to quantify interdigitation we measured the density profiles of
the alkane chains for the two leaflets separately.
Figure~\ref{fig:interdig} shows this data for three selected
concentrations. At low concentrations there is no appreciable
difference to the small interdigitation existing already in the
pure bilayer. However, if we increase the concentration to about
2\% mol of n--butanol we see clearly increased interdigitation.
This indicates that that low butanol concentration has little
impact on lipid structure. However, as the concentration increases
the bilayer structure is massively changed but not abandoned.

\begin{figure}
\includegraphics[width=5.5cm]{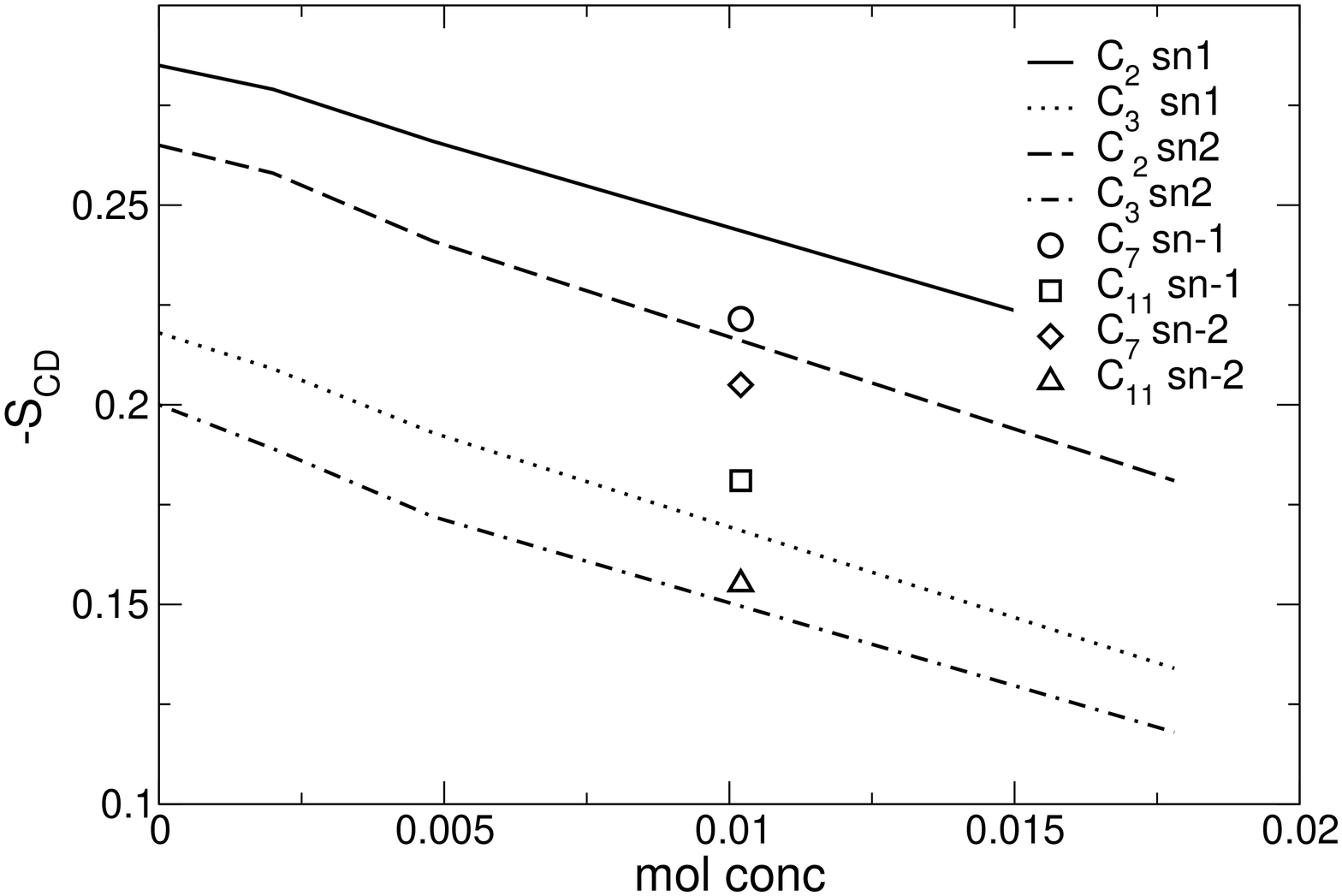}\hspace{1cm}
\includegraphics[width=5.5cm]{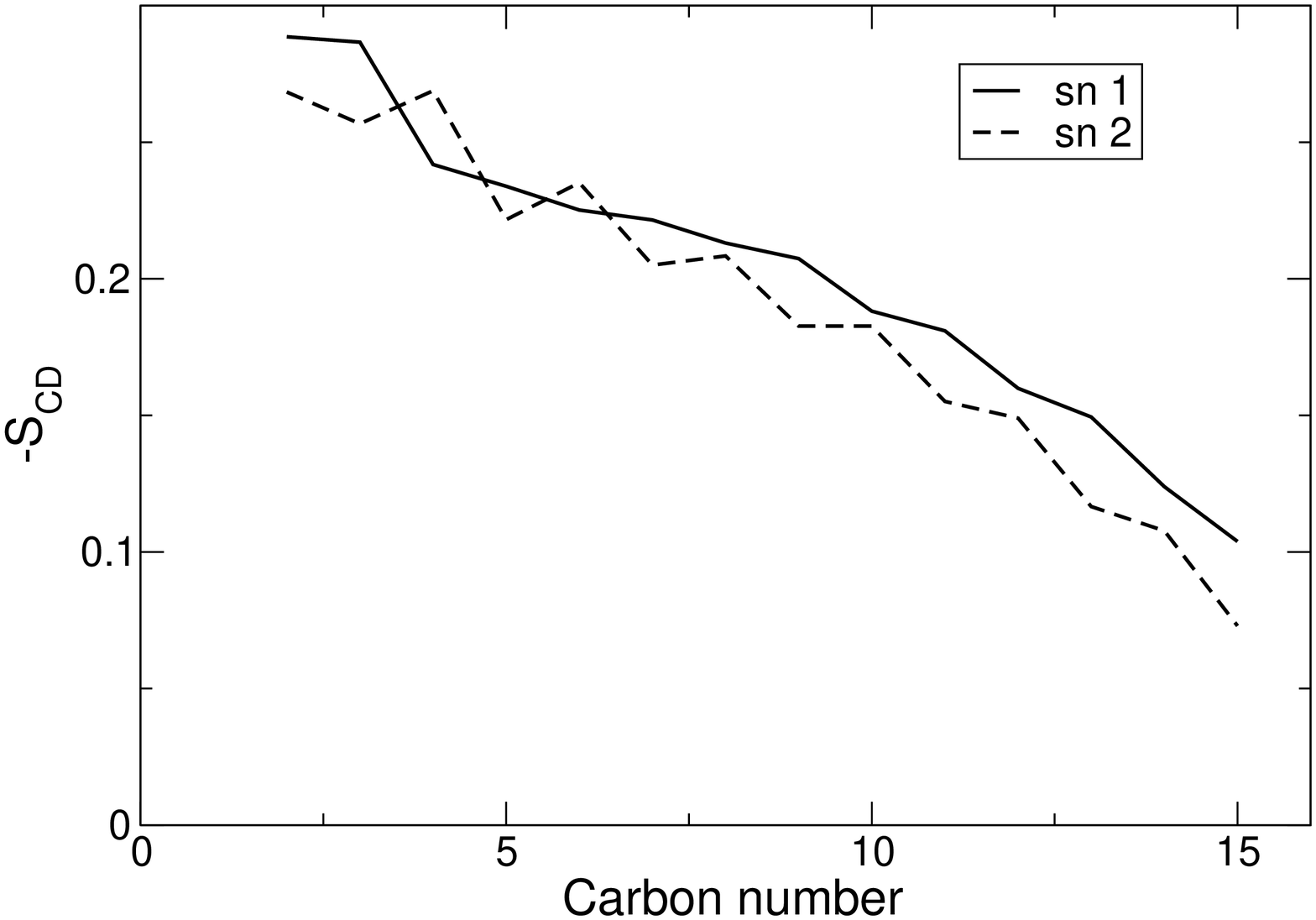}
\caption{The left hand panel shows the dependence of the deuterium
order parameter on alcohol concentration for coarse--grained and
atomistic simulations. The atom numbers in the coarse--grained
case represent the interaction sites with numbering starting at
the head group. The right hand panel shows the deuterium order
parameter for the individual lipid chains from the atomistic
simulation at a molar n--butanol concentration of 0.0102.}
\label{fig:order}
\end{figure}

The bilayer structure is often characterized using the deuterium
order parameter which is defined as
\begin{eqnarray}
  - S_{CD}&=&\frac{2}{3}S_{xx} + \frac{1}{3}S_{yy},\\
  S_{\alpha\beta}&=&\langle3\cos\Theta_{\alpha}\cos\Theta_{\beta}-
  \delta_{\alpha\beta}\rangle,\qquad\alpha,\beta=x,y,z\\
  \cos\Theta_{\alpha}&=&\hat{e}_{\alpha}\hat{e}_z,
\end{eqnarray}
where $\hat{e}_z$ is a unit vector in the laboratory
$z$--direction and $\hat{e}_{\alpha}$ is a unit vector in the
local coordinate system of the tails. The order parameter
calculation is based on three connected carbons (or
coarse--grained interaction sites) C$_{i-1}$,C$_{i}$, and
C$_{i+1}$ and $\vec{e}=\vec{r}_{i+1}-\vec{r}_{i-1}$.
Figure~\ref{fig:order} shows this order parameter. Since the
coarse--grained chains are composed of four interaction centers,
only two order parameters per chain can be defined. We see a
slight difference in the order between the $sn-1$ and $sn-2$
chains where the $sn-1$ chain is the alkane tail directly
connected to the headgroup and $sn-2$ is separated from the head
group by the glycerol backbone.

With increasing concentration the order parameter decreases
significantly at all positions along the chains. This is shown in the
left--hand panel of Figure~\ref{fig:order}. The effect is independent
of the distance from the headgroup and the identity of the chain
($sn-1$ or $sn-2$) as all the 4 curves in Figure~\ref{fig:order} are
parallel. This indicates that the alcohol not only influences the
headgroup and its immediate vicinity but the whole layer. It is
interesting to note that the coarse--grained model can distinguish
between the two chains of the lipid, it actually overestimates the
difference. The $sn-1$ chain is more highly ordered in all cases. In
order to directly compare the data between atomistic and coarse
grained we assume that the four interaction sites in the lipid tails
represent the four quarters of the DPPC tail not taking the glycerols
into account. This means we take the middle atom of the second and
third quarter of the atomistic tails (C$_7$ and C$_11$) to compare,
these are shown as symbols in the left hand panel of figure~\ref{fig:order}.  
We see that for both atoms we get satisfactory agreement. Due to the
intrinsic averaging of the coarse--grained model the difference
between the models is about 10\%. We could obtain better agreement if
we used other atomistic carbons to directly compare. However, that
would not be in the spirit of the used mapping. The right hand panel
additionally shows the full atomistic order parameter for the case of
the 0.0105 concentration (i.e. 39 alcohols).

\section{Conclusion}
In this study, we examined the effects of alcohol concentration on
a lipid bilayer and showed that a coarse--grained model enables a
wider range of parameters to be surveyed than is possible using
only atomistic models. The accuracy of the coarse-grained model
was examined by comparing experimentally important quantities with
atomistic simulations and experimental data  in a relatively short
amount of computer time. This comparison indicated which
coarse-grained parameters need to be refined.

The two models that we are comparing represent fundamentally the
same system and the combination of the two can be used to
investigate various bilayer properties. However, there are clear
differences and a coarse--grained model should only be used with
caution and in conjunction with an atomistic model as a basis. In
this study we mainly see that the alcohol affinity to the actual
interface is overestimated in the coarse--grained model.
Nonetheless, most of the relevant properties are very well in
agreement between the two models. We will continue evaluating
bilayer properties for both models. From these studies, we will be
able to identify important parameters that should be incorporated
into future coarse-grained models. For the results here the exact
coarse grained interactions were not very important as we
performed additional simulations with slightly changed parameters
and obtained the same conclusions.

We showed that increasing alcohol concentration leads to a linear
increase of the area per molecule and to a linear decrease in the
chain order parameter. The alcohols are modeled as amphiphilic
molecules and tend to reside at the interface, i.e. between the
headgroup and the tails. Butanols are more likely to enter the
tail region shorter alcohols. At high butanol concentrations, the
interface can become saturated  and cause butanol molecules to
return to the water phase. Lipid conformational changes increase
with alcohol concentration and this corresponds to an increase in
leaflet interdigitation.
\bibliography{standard}
\bibliographystyle{JPSBnew}
\end{document}